\documentclass[floatfix,prl,twocolumn,superscriptaddress,showpacs]{revtex4}
\usepackage{amsmath}
\usepackage{amsfonts}
\usepackage{amssymb}
\usepackage{graphicx}
\usepackage{color}

\newcommand{\km}{\left<k\right>}

\begin{document}

\author{Salvatore Mandr\`a}
\affiliation{Universit\`a degli Studi di Milano, Dip.
    Fisica, Via Celoria 16, 20133 Milano, Italy } 
\affiliation{INFN, Sezione di Milano, Via Celoria 16, 20133 Milano, Italy}

\author{Santo Fortunato}
\affiliation{Complex Networks Lagrange Laboratory, ISI Foundation, Torino, Italy}

\author{Claudio Castellano}
\affiliation{SMC, INFM-CNR and
Dipartimento di Fisica, ``Sapienza'' Universit\`a di Roma,
P.le Aldo Moro 2, I-00185 Roma, Italy}

\title{Coevolution of Glauber-like Ising dynamics and topology}

\begin{abstract}
We study the coevolution of a generalized Glauber dynamics for Ising spins, with tunable threshold,
and of the graph topology where the dynamics takes place. 
This simple coevolution dynamics generates a rich phase diagram in the space of the two parameters of the model,
the threshold and the rewiring probability. The diagram displays
phase transitions of different types: spin ordering, percolation, connectedness. 
At variance with traditional 
coevolution models, in which all spins of each connected component of the graph have equal value
in the stationary state, we find that, for suitable choices of the parameters, the system may 
converge to a state in which spins of opposite sign coexist in the same component, organized in 
compact clusters of like-signed spins. Mean field calculations enable one to estimate some features of the phase diagram.

\end{abstract}

\def\mean#1{\left\langle#1\right\rangle}
\def\part#1{\left(#1\right)}
\def\parq#1{\left[#1\right]}

\maketitle

In recent times there has been an increasing attention, by the statistical
physics community, towards applications to social systems and relative
phenomena~\cite{castellano09}. The goal is the description and possibly 
the prediction of collective features of processes involving large numbers
of individuals without detailed information on the characteristics of
the single elements, much like it happens in the physics of phase
transitions~\cite{binney}.
Many simple models have been devised, inspired by intuitive ideas on 
how social interactions between individuals take place.
Such models are often variations of known models of statistical physics,
or entirely new and interesting types of dynamics. The main ingredients
are a graph, representing the social network of interactions (acquaintances)
between individuals, and a set of local rules, indicating how the state of an
agent is affected by (or affects) the state of its neighbors.
The graph may be a lattice or have a more complex topology, reflecting 
properties observed in real social
networks~\cite{Newman:2003,vitorep,barratbook}.
Usually one studies the model dynamics on a given graph topology,
which remains frozen during the whole evolution of the process.
However, in real social phenomena the dynamics of states
is often coupled to the transformation of the social network where
the process takes
place, as the network evolves as well, and the time scales of the two
evolutions may be comparable. So, a realistic description of social processes
must consider the coevolution of state dynamics and network topology.
In the last years several coevolution models have been
proposed~\cite{zimmermann04,Ehrhardt06,Holme06,gil06,grabowski06,centola07,vazquez07,vazquez08,nardini08,kozma08,benczik08,klimek08}.
The interaction rules of such models combine both changes in the states of
the agents and in the link structure of the underlying graph.
Frozen states of the dynamics are usually characterized by a network composed of one or more connected components 
with all agents in each component being in the same state.
Indeed, the dynamics of states
in each component becomes independent of the dynamics ruling the states of the other components and, while the agents of each component converge to the
same state, such a state will usually be different
from one component to another. 
In several models both scenarios, i. e. one component with all agents in the same state and 
two or more separate components each in a different state, can be reached by suitable choices
of the parameter weighing the relative importance of the dynamics of the
states versus that of the graph
topology~\cite{Holme06,gil06,centola07,vazquez08,nardini08,kozma08}.
Such a scenario is quite simple but it is not very realistic. For example,
these models cannot describe the situation in which different groups of people
sharing the same state (domains) coexist in the same component, something which is likely to happen in society.
In this letter, we present the first model that accounts for this situation
as well. 
Our model is based on a simple Glauber-type dynamics for Ising
spins~\cite{glauber63}.
It can also be seen as a sort of threshold model~\cite{granovetter78}
where disorder is in the topology and not in the thresholds.
We show that, in spite of its simplicity, the model has a very rich
behavior, with several phases, separated by transitions involving both
the spin states and the graph topology.   

The starting point is a random graph \`a la Erd\"os-R\'enyi~\cite{erdos}
with $N$ nodes and $M$ links, with
$M = \frac{\left<k\right> N}{2}$, $\left<k\right>$ being the average
degree of the graph. 
We stress that the main results do not depend on the initial network topology because
the rewiring dynamics leads inevitably to a random network with a Poisson degree distribution. 
Agents lie on the nodes of the graph, and are endowed with binary
states (spins) $\sigma=\pm 1$, which are initially assigned at random with
equal probability $1/2$.
The dynamics of the model is defined by iterating the following
update rule:
\begin{enumerate}
\item A node $i$ is selected at random: we indicate with $k_i$ the number
of its neighbors and with $l_i$ the number of neighbors in the same state.
\item If $l_i/k_i \geq s$, where $0\leq s\leq 1$, the node is stable and nothing happens;
otherwise a neighbor $j$ with $\sigma_j \not= \sigma_i$ is randomly chosen and
\begin{itemize}
\item with probability $\phi$, $i$ cuts its link to $j$ and attaches
it to a randomly chosen node $l$ such that $\sigma_l = \sigma_i$ and $l$
is not already connected to $i$~\footnote{If no such node exists no action
is taken.} (\emph{rewiring});
\item with probability $1-\phi$, $i$ adopts $j$'s state (\emph{spin flip}).
\end{itemize}
\end{enumerate}
The model has two relevant parameters: the threshold $s$ and the probability
of rewiring $\phi$. 
The threshold sets the minimum fraction of neighbors in the same state that
a node must have to be stable. In this respect it is a measure of the
sensitivity of agents against the social pressure exterted by
neighbors with opposite state.
If $s$ is very small virtually all nodes are stable, i. e. they do not flip
their spin nor rewire their connections.
On the contrary, if $s$ is close to 1 only nodes fully surrounded by nodes
in the same state are stable.
When $s=1/2$, the spin dynamics is essentially the Glauber
dynamics of Ising spins at zero temperature.
For $s<1/2$ the dynamics is rather uninteresting, 
as already in the initial condition typically nodes have at least half of the
neighbors in the same state and hence they are stable.
Therefore we focus on the range $1/2\leq s\leq 1$.
In general, the presence of a threshold allows for the existence of
{\it unsatisfied links}  (i.e. links joining agents with different states) 
in stable states of the system, at variance with standard coevolution models.
The rewiring probability $\phi$ is a measure of the relative importance
of the rate at which the network evolves with respect to the rate at which
the state of the node changes. The extreme values correspond to pure spin
dynamics on a fixed network topology ($\phi = 0$), and to
pure network evolution, with no spin dynamics ($\phi = 1$). 

The phase diagram of the model has a remarkably rich structure.
To study its features, we monitor the behavior of some standard observables,
the magnetization $m=\sum_i\sigma_i/N$ and the density of unsatisfied links
$\rho= \sum_{i<j} A_{ij}[1-\delta(\sigma_i,\sigma_j)]/M$,
where $A_{ij}$ is the element of the adjacency matrix of the graph
($A_{ij}=1$, if $i$ and $j$ are neighbors, otherwise $A_{ij}=0$), and 
$\delta$ is Kronecker's delta function. 
Moreover we consider the convergence time $t_c$, defined as the time needed
to reach a frozen configuration, at which the dynamics stops
(i.e. $l_i/k_i>s$ for any $i$).
\begin{figure*}
  \centering
  \includegraphics[width=0.34\textwidth]{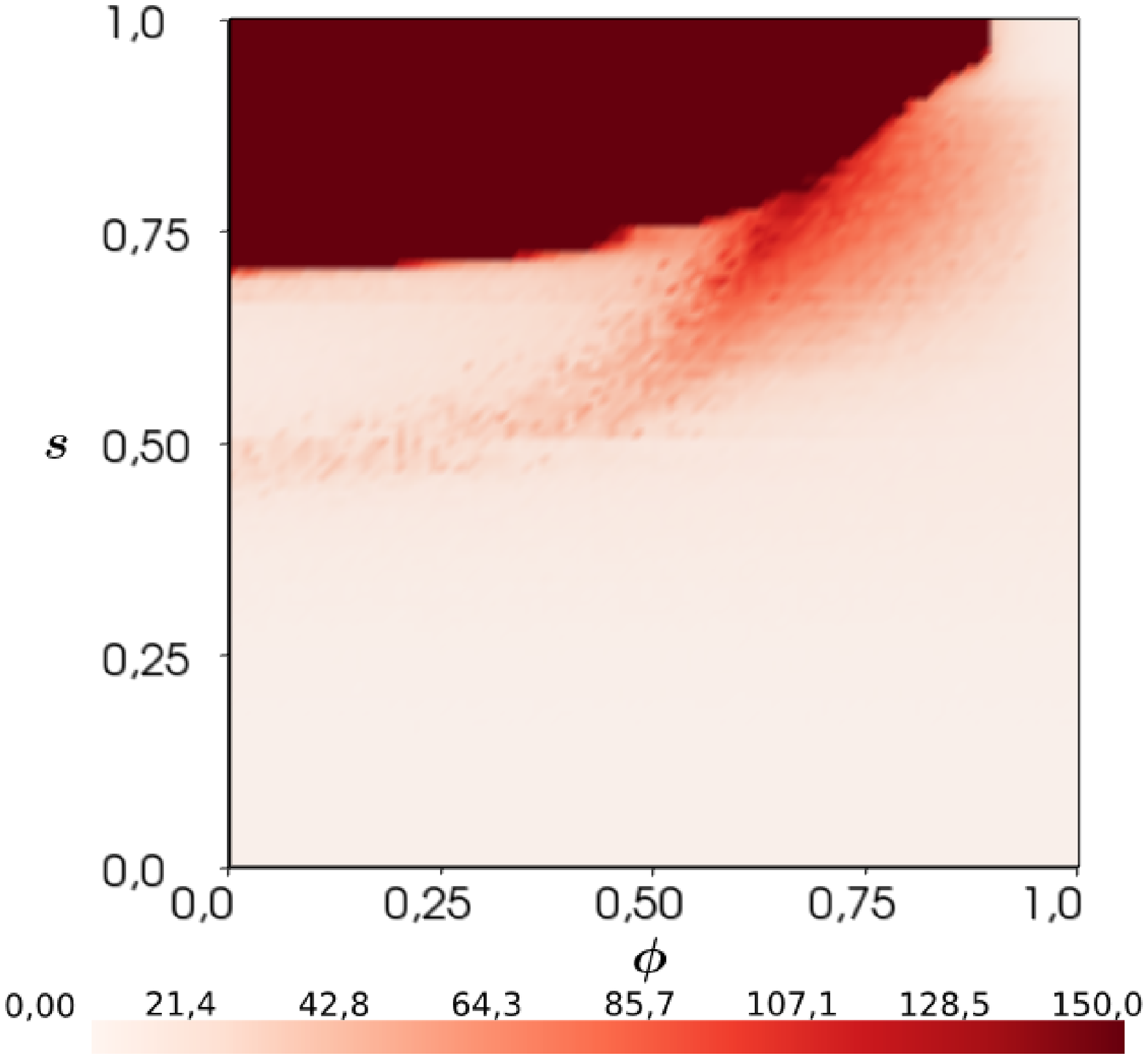}
  \includegraphics[width=0.35\textwidth]{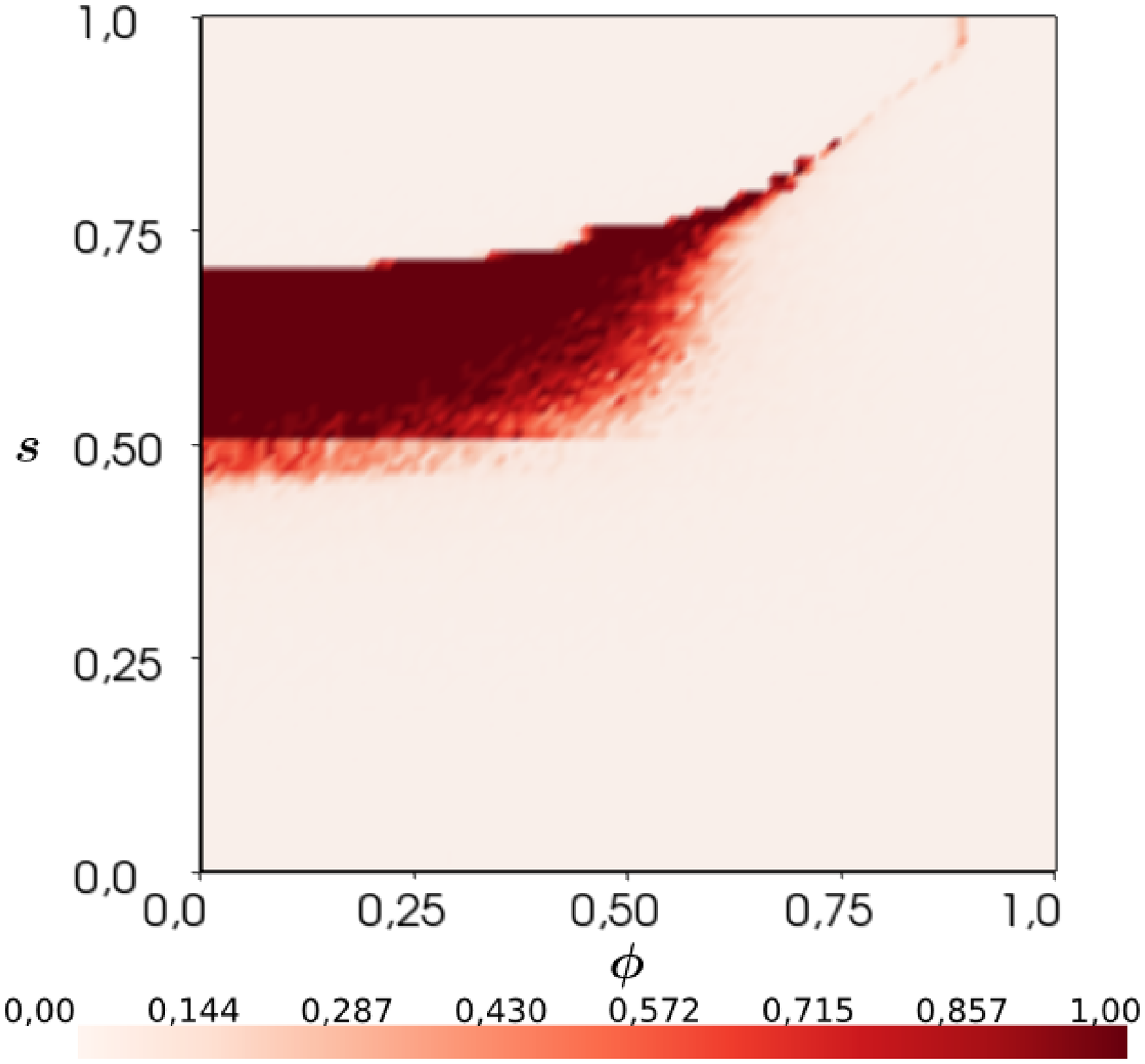}\\
  \vspace{0.5cm}
  \includegraphics[width=0.35\textwidth]{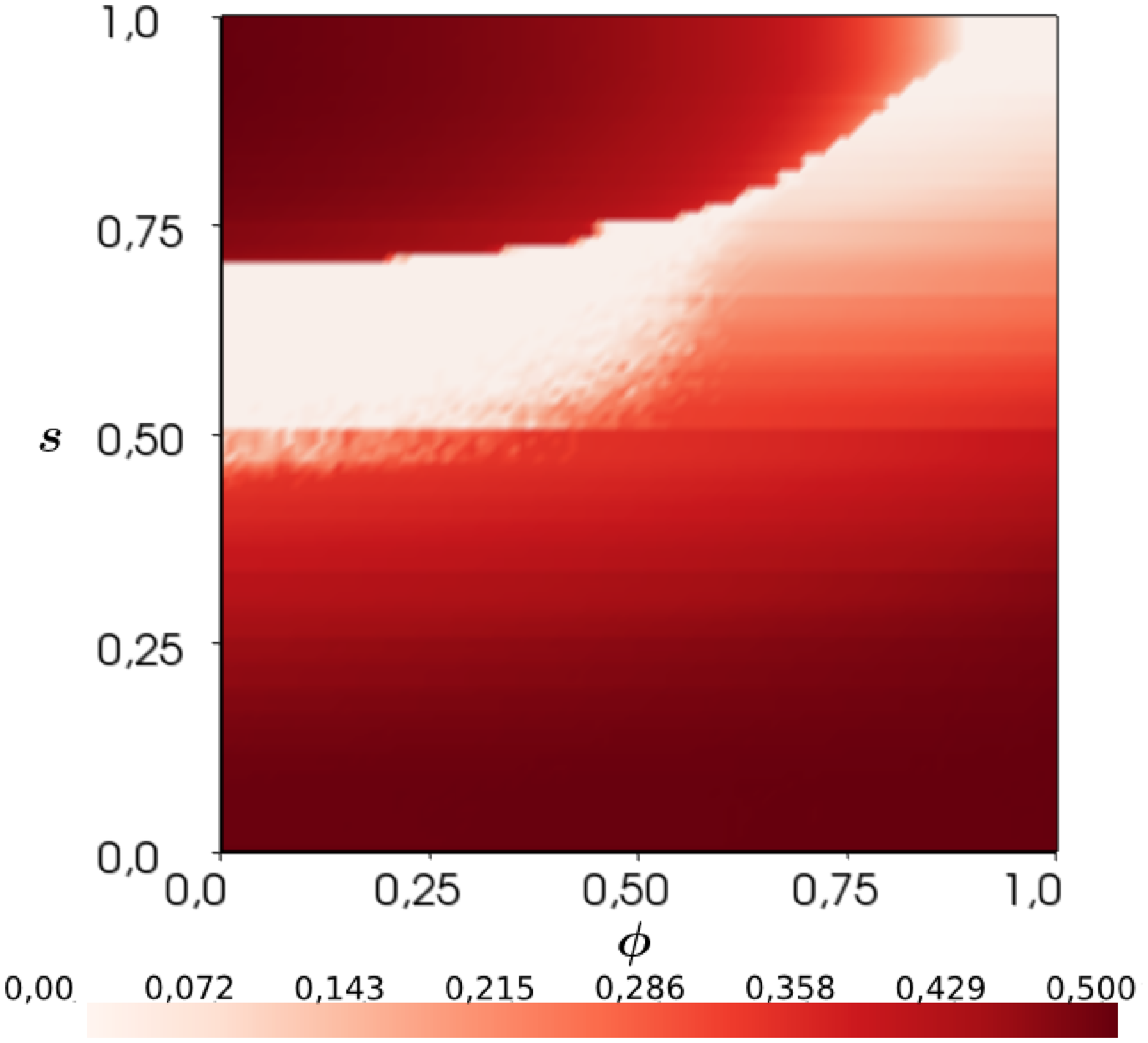}
  \includegraphics[width=0.35\textwidth]{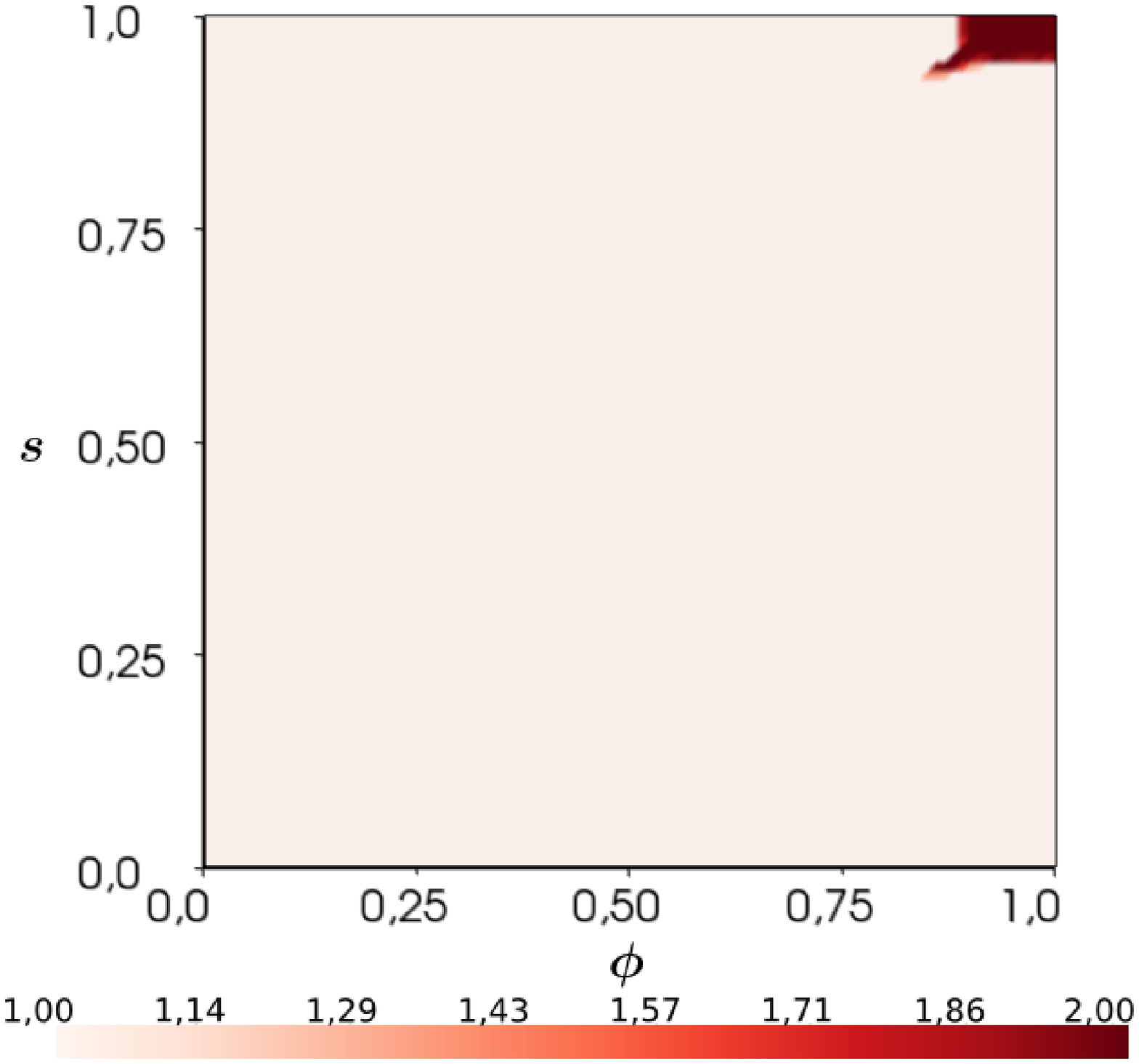}
  \caption{(color online) 
Dependence on the threshold $s$ and the rewiring probability
$\phi$ of several variables: the convergence time $t_c$ (top left), the absolute value
of the magnetization (top right), the density of unsatisfied links
(bottom left) and the number of connected components (bottom right).
Data are obtained from numerical simulations on a graph with $N = 50000$, 
$\left<k\right> = 10$. The darker color in the top left panel means that
the convergence time is larger than $150$ times the number of nodes.
Data in the three other panels are computed after
\mbox{$1.5\times 10^6$} iterations.}
\label{fig1}
\end{figure*}
In Fig.~\ref{fig1} we report in the plane $(s,\phi)$
the numerical value of $t_c$, as well as the value of $|m|$ and $\rho$ and
the number of connected components of the graph,
after a very long run, sufficient to reach the stationary state.
From the behavior of the convergence time, it is clear that the parameter
space is divided into two regions. In the upper left zone (denoted as $S-0$)
there is a phase with ongoing dynamic activity.
In this region the convergence time $t_c$ diverges exponentially with $N$,
so that it is effectively infinite for systems of any reasonable size.
However, the system reaches a stationary state, with constant value of the
observables.
Elsewhere, instead, the dynamics leads in a finite time to an absorbing frozen state, with no dynamics.
The two regions are separated by an absorbing-state phase
transition~\cite{Marrobook}. 
The values of $|m|$ and $\rho$ indicate that the active phase is disordered:
the average magnetization remains zero and the density of unsatisfied links
remains high. In this phase, due to the high value of $s$, sites are rarely
stable and they keep rewiring, looking for similar
partners, as in Refs.~\cite{centola07} and~\cite{vazquez07}.
The structure of the absorbing phase is much richer, as one can identify
several distinct subphases, with various types of internal organization.
For small values of $\phi$ and $s$, there is a phase (A-I) where $|m|=1$
and $\rho=0$. Here the relatively slow rewiring process allows spin ordering
to be completed while the topology remains globally connected.
The opposite occurs in the upper right corner of the parameter space
(phase A-II).
For large $s$ and $\phi$, the value of $\rho$ is the same
of the A-I zone, but in the A-II zone the number of connected
components is $2$: the system splits into two topologically separated sets of
similar size, one of them fully ordered with $\sigma=1$ and the other
with $\sigma=-1$.
Phases A-I and A-II correspond to those found in the voter
model~\cite{vazquez08}.
To understand the organization of the system in the rest of the plane,
we study the presence and extent of homogeneous domains, intended
as subsets of nodes of the network with two properties: 1) all nodes of
the subset are in the same state and
2) any pair of nodes of the subset can be joined with a path
within the subset.
We then measure two new observables: the number of homogeneous domains
in the network and the relative size of the largest of them. 
In Fig.~\ref{figncfc} we plot these two observables as a function of $s$
for a fixed large value of the rewiring probability $\phi$.
It is possible to identify two new phases, delimited by two threshold values 
$s_p$ and $s_q$ (indicated by the two vertical dotted lines).

For $s<s_p$ there are just microscopic domains and their number is
proportional to the number of nodes in the network. In this phase ($A-0$), 
that spans the whole range of $\phi$ for $s<1/2$, $m=0$ and $\rho\sim 1/2$. Stability is rather
easy to reach for all nodes, after few spin flips or link rewirings.
We stress that $A-0$ and $S-0$ are different: 
in the case of $A-0$ an absorbing state is always reached,
while for $S-0$ the system reaches only a dynamic stationary state.
At $s=s_p$ a percolation transition~\cite{staufferbook} takes place:
for $s_p<s<s_q$ the dynamics lasts long enough to allow for the formation
of macroscopic domains (typically two) that grow bigger as $s$ is increased
up to the point where they occupy the whole system (phase $A-D$).
Finally, for still higher values of the threshold $s>s_q$ the macroscopic
domains become topologically disconnected from each other
and coincide with the two connected components of the network (phase $A-II$).
\begin{figure}
  \centering
  \includegraphics[angle=-90, width=\columnwidth]{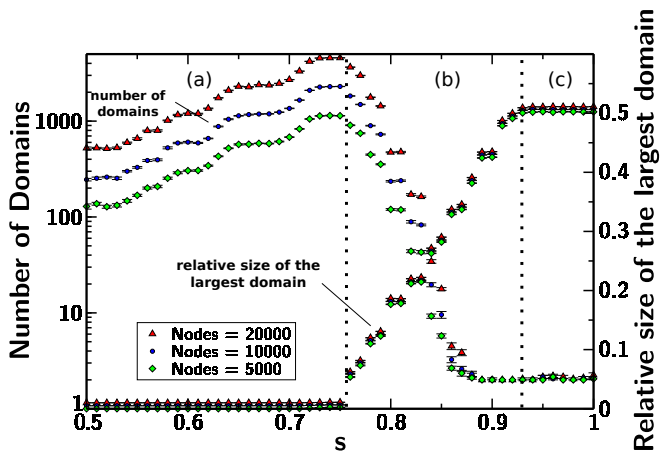}
  \caption{(color online)
Number of domains and density of the largest domain as a function of
$s$, obtained from simulations on a graph with $\left<k\right> = 10$
and $\phi = 0.95$. 
}
\label{figncfc}
\end{figure}
Based on this evidence, we schematically represent in Fig.~\ref{figdiag}
the phase-diagram of the model.
Our simulations, performed up to size $N=50000$, seem to indicate that
the parameter space is divided into genuine phases separated by
well-defined transition lines. However, a detailed investigation 
of the nature of all of them (and of the associated critical behavior)
is numerically very demanding and goes beyond the scope of the present paper.\\
\begin{figure}
  \centering
  \includegraphics[angle=-90,width=\columnwidth]{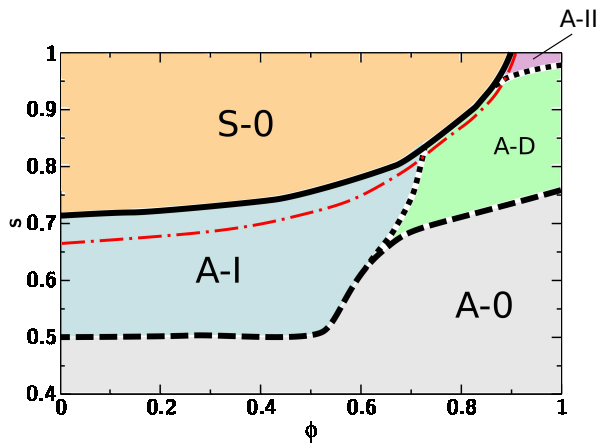}
  \caption{(color online) Schematic representation of the phase diagram
obtained from numerical simulations on a graph with
$N = 50000$, $\left<k\right> = 10$, after \mbox{$1.5\times 10^6$} iterations. 
We can identify two types of phases: the dynamic stationary phase
(denoted by $S-0$) and four absorbing phases 
(denoted by $A$), divided by the black solid line.
Each absorbing phase is characterized by a different organization of the
network in domains (see text for details).
The (red) dot-dashed line corresponds to the numerical solution of mean-field
equations.}
\label{figdiag}
\end{figure}

Some of the features of the phase-diagram are recovered
(see Fig.~\ref{figdiag}) via a mean field (MF) approach,
similar to the one in Ref.~\cite{vazquez08}.
At each step, a node with $k$ links is randomly chosen. 
Since the rewiring
dynamics leads to a network with a Poisson degree distribution,
the probability $P_k(t)$ to extract a node with $k$ links
is supposed to be a Poisson distribution with mean $\km$ at each step.
Denote with $j$ the number of unsatisfied links and with $\mathcal{P}_{k,j}$
the probability that the chosen node is not stable and hence must be
updated.
With probability $\phi\,\mathcal{P}_{k,j}$ a random unsatisfied link is
rewired. In this case, the density of unsatisfied links
changes by $\Delta\rho = -\frac{2}{\km\! N}$. On the contrary,
with probability $\part{1-\phi}\,\mathcal{P}_{k,j}$ the state of the
node is flipped and the density of unsatisfied links changes by the quantity
$\Delta\rho = \frac{2\part{k-2j}}{\km\! N}$.
Using these expressions, it is possible to write the
time evolution master equation for a generic update rule
\begin{align}\label{eq:master}
	\frac{d\rho}{dt} &= 
		\sum_k \frac{P_k}{1/N}\sum_{j = 0}^k 
		B_{k,j}^{(\rho)} \,\mathcal{P}_{k,j}
		\parq{(1-\phi)\frac{2\part{k-2j}}{\km\! N} - \phi\frac{2}{\km\! N}}\nonumber\\
		&= F_{\phi,s}(\rho),
\end{align}
where $1/N$ is the temporal interval between successive steps
and $B_{k,j}^{(\rho)}$ is the probability to find a node with $j$ unsatisfied
links. In a MF spirit, all nodes of the network can be considered equivalent 
and the probability to have an unsatisfied link can depend only on the global
observable $\rho$.
Thus, the probability to have an unsatisfied link is taken independent
for each node and is well approximated by a binomial distribution
$B_{k,j}^{(\rho)} = \binom{k}{j}\rho^j(1-\rho)^{k-j}$.\\

In the case of the voter model
the update probability is simply $j/k$, and it is possible to analytically
solve it~\cite{vazquez08}. In our model, the update probability is
$\mathcal{P}_{k,j} = \theta\part{j/k + s -1}$ where 
$\theta\part{x}$ is the Heaviside step function. Due to the nonlinearity of
Eq.~(\ref{eq:master}) an analytical expression can be found
only for $s = 1$. In such a case the right-hand side of Eq.~\ref{eq:master}
has the simple expression
\begin{align}
	F_{\phi,s}(\rho) &= 2(1-\phi)(1-2\rho)-\frac{2}{\km}\phi\nonumber \\
	&-e^{-\km\rho}\part{2(1-\phi)(1-\rho)-\frac{2}{\km}\phi}.
\end{align}
For $\rho_s$ sufficiently small, the stationary solution of Eq.~(\ref{eq:master}) has then the form
$\rho_s = \frac{2\parq{\km\part{1-\phi}-1}}{\km\parq{2+\km(1-\phi)-3\phi}}$.
For $\phi < \phi_c(s = 1) = \frac{\km - 1}{\km}$
there is an active stationary state with $\rho_s > 0$. For larger
values of $\phi$ the density of unsatisfied links is zero, corresponding
to an absorbing phase. The transition is predicted to be continuous.
For $\km = 10$ the critical value is $\phi_c(s = 1) = 0.9$,
in agreement with numerical simulations.
It is possible to determine numerically the transition line $\phi_c(s)$ for
any value of $s$. The resulting curve is reported in Fig.~\ref{figdiag}.

The simple model we have proposed offers a surprisingly rich variety of possible scenarios, 
by varying the two parameters $s$ and $\phi$. In particular, phase boundaries correspond to 
magnetization, connectedness and/or percolation transitions.
The most striking feature, absent in all other models
of coevolution, is the existence of a phase where stable homogeneous domains coexist in the system,
even if the latter is not split into components. This feature is due to the presence of the threshold $s$:
models characterized by a threshold are likely to display this type of behavior and represent a promising
option for a realistic description of social phenomena. We stress however that the
goal of this paper was not a description of a specific real world phenomenon,
but rather the investigation of what are the possible qualitative outcomes
when threshold dynamics and rewiring operate simultaneously.


\begin{thebibliography}{20}

\bibitem{castellano09} C. Castellano, S. Fortunato and V. Loreto, Rev. Mod. Phys. {\bf 81}, 591 (2009).

\bibitem{binney} J. J. Binney, N. J. Dowrick, A. J. Fisher and M. E. J. Newman, {\it The Theory
of Critical Phenomena}, Oxford University Press, Oxford, UK (1992).

\bibitem{Newman:2003} M.~E.~J. Newman,
SIAM Review {\bf 45}, 167 (2003). 

\bibitem{vitorep} 
S. Boccaletti, V. Latora, Y. Moreno, M. Chavez and D.-U. Hwang,
Phys. Rep. {\bf 424}, 175 (2006).

\bibitem{barratbook}
A. Barrat, M. Barth\'elemy and A. Vespignani, {\it Dynamical Processes on Complex Networks}, 
Cambridge University Press, Cambridge (2008).

\bibitem{zimmermann04} M.~G. Zimmermann, V.~M. Egu\'{\i}luz, and M. San Miguel,
Phys. Rev. E {\bf 69}, 065102(R) (2004).

\bibitem{Ehrhardt06} G.~C. Ehrhardt, M. Marsili, and F.
Vega-Redondo, Phys. Rev. E {\bf 74}, 036106 (2006).

\bibitem{Holme06} P. Holme and M.~E.~J. Newman, Phys. Rev. E {\bf 74}, 056108
(2006).

\bibitem{gil06} S. Gil and D.~H. Zanette, Physica D {\bf 224},  156  (2006).

\bibitem{grabowski06} A. Grabowski and R.~A.Kosi\'nski, Phys. Rev. E {\bf 73}, 016135 (2006).

\bibitem{centola07} D. Centola, J.~C. Gonz\'alez-Avella, V.M. Egu\'{\i}luz, and
M. San Miguel, J. of Conflict Resol. {\bf 51}, 905 (2007).

\bibitem{vazquez07} F. Vazquez, J. C. Gonzalez-Avella, V. M. Egu\'{\i}luz, and
M. San Miguel, Phys. Rev. E {\bf 76}, 046120 (2007).

\bibitem{vazquez08} F. Vazquez, V.~M. Egu\'iluz and M. San Miguel, Phys. Rev. Lett. {\bf 100}, 108702 (2008). 

\bibitem{nardini08} C. Nardini, B. Kozma, and A. Barrat, Phys. Rev. Lett. {\bf 100}, 158701 (2008).

\bibitem{kozma08}
B. Kozma and A. Barrat, Phys. Rev. E {\bf 77}, 016102 (2008).

\bibitem{benczik08}
I. J. Benczik, S. Z. Benczik, B. Schmittman, and R. K. P. Zia,
Europhys. Lett. {\bf 82}, 48006 (2008).

\bibitem{klimek08}
P. Klimek, R. Lambiotte, and S. Thurner,
Europhys. Lett. {\bf 82}, 28008 (2008).

\bibitem{glauber63} R.~J. Glauber, J. Math. Phys. {\bf 4}, 294 (1963).

\bibitem{granovetter78}
M. Granovetter,
Am. J. Sociol. {\bf 83}, 1420 (1978).

\bibitem{erdos} P. Erd\"os and A. R\'enyi,
Publ. Math. Debrecen {\bf 6}, 290 (1959).

\bibitem{Marrobook} J. Marro and R. Dickman, {\em Nonequilibrium
 phase transitions in lattice models} (Cambridge University Press, Cambridge, 1999).

\bibitem{staufferbook} D. Stauffer and A. Aharony, {\it Introduction to Percolation Theory}, Taylor {\&} Francis, London (1994).

\end{thebibliography}
\end{document}